\PassOptionsToPackage{table}{xcolor}
\documentclass[sigconf]{acmart}

\usepackage{kotex}
\usepackage{multirow}
\usepackage{booktabs}
\usepackage{colortbl}
\usepackage{tabularx}
\usepackage{array}
\usepackage{makecell}

\definecolor{lightgray}{rgb}{0.95,0.95,0.95}
\AtBeginDocument{%
  \providecommand\BibTeX{{%
    \normalfont B\kern-0.5em{\scshape i\kern-0.25em b}\kern-0.8em\TeX}}}

\copyrightyear{2026}
\acmYear{2026}
\setcopyright{cc}
\setcctype{by}
\acmConference[DIS '26]{Designing Interactive Systems Conference}{June 13--17, 2026}{Singapore, Singapore}
\acmBooktitle{Designing Interactive Systems Conference (DIS '26), June 13--17, 2026, Singapore, Singapore}
\acmDOI{10.1145/3800645.3812920}
\acmISBN{979-8-4007-2563-0/2026/06}

\begin{document}

\title[Value-Sensitive AI for Prayer]{Value-Sensitive AI for Prayer: Balancing the Agencies Between Human and AI Agents in Spiritual Context}

\author{Soonho Kwon}
\email{soonho@gatech.edu}
\orcid{0000-0002-2783-6364}
\affiliation{%
  \institution{Georgia Institute of Technology, \linebreak School of Interactive Computing}
  \city{Atlanta}
  \state{GA}
  \country{USA}
  \postcode{30332}}

\author{Dong Whi Yoo}
\email{dy22@iu.edu}
\orcid{-}
\affiliation{%
  \institution{Indiana University Indianapolis, \linebreak Luddy School of Informatics, Computing, and Engineering}
  \city{Indianapolis}
  \state{IN}
  \country{USA}
  \postcode{46202}}

\author{Shaowen Bardzell}
\email{shaowen@cc.gatech.edu}
\orcid{-}
\affiliation{%
  \institution{Georgia Institute of Technology, \linebreak School of Interactive Computing}
  \city{Atlanta}
  \state{GA}
  \country{USA}
  \postcode{30332}}

\author{Younah Kang}
 \email{yakang@yonsei.ac.kr}
 \affiliation{%
   \institution{Yonsei University, \linebreak Information and Interaction Design}
   \city{Seoul}
   \country{Republic of Korea}
   \postcode{03722}}

\renewcommand{\shortauthors}{Kwon et al.}

\begin{abstract}
How could AI enter a deeply value-laden realm of human lives? Drawing on key values and practices associated with praying identified through a diary study, we presented our participants with four speculative, conceptual value-sensitive AI systems to “assist” prayer practices. The conceptual designs served  as provocations to co-reflect on how AI interventions might shape their praying experiences. Our findings suggest that a sense of authenticity (or a genuine connection to the divine) is a crucial value, while the mere presence of AI was often perceived as diminishing this authenticity, particularly when AI assumed too much agency in guiding prayer practices. Based on our findings, we argue for the importance of AI agent designs that recognize users’ agency in shaping their interaction with AI. We further explore how this may be possible by leveraging interpretive openness, perhaps through AI’s inexplicability as a resource for personal meaning-making, and by recognizing non-use of AI as a legitimate design choice.

\end{abstract}




\maketitle


\section{Introduction}
With the increasing popularity of commercial AI, attempts, hype, and concerns surrounding its implementation in spiritual domains are growing \cite{smith2024un,smith2026understanding}. Namely, in South Korea, the generative AI service \textit{Chowon} has gained significant popularity among Christians. This service utilizes AI to address users' religious concerns, providing personalized Bible verses or AI-generated interpretations. Integrating AI into spirituality and religion prompts design and HCI researchers to reflect on its implications, critically exploring how it might shape both the landscape of techno-spirituality and human-AI interaction.

The concept of techno-spirituality, in which spirituality is supported and mediated by technology, and technologies are designed to enrich spiritual practices, has been an important topic in HCI~\cite{buie2019let}. One of the biggest significances of techno-spirituality for the HCI and design community lies in its expansion of system goals beyond the workplace-oriented goals such as efficiency and speed, closely aligning with ``third-wave HCI'' movements \cite{bodker2006second}. Introducing the concept, \citeauthor{bell2006no} highlights how techno-spirituality enables consideration of diverse and nuanced values such as grace, humility, privacy, and identity~\cite{bell2006no}. Additionally, multiple studies underscore the inherent value-oriented nature of spiritual activities~\cite{laird2004measuring, whittington2010prayer, poloma1991effects}, emphasizing the importance of designing for deeper values rather than prioritizing speed and efficiency in spiritual contexts~\cite{buie2013spirituality, wolf2022spirituality, wolf2023designing}. Building on these insights, we consider using AI for spiritual purposes a promising avenue for envisioning alternative applications of AI that transcend the goals of efficiency, accuracy, and productivity.

HCI researchers have a long history of understanding and supporting diverse values in technology design. Notably, the concept of value-sensitive design (VSD) emphasizes that the diverse values of specific stakeholders must be considered throughout the entire design process~\cite{friedman1996value}. Since its inception, value-sensitive design has evolved to emphasize the importance of situated realities, where specific user values are derived from their lived experiences, rather than imposing pre-defined values. \cite{LeDantec2009value, borning2012next}. While early VSD efforts focused on values that are often discussed in system design, such as benevolence~\cite{friedman1996value}, researchers have also emphasized the situated and contextual nature of values \cite{borning2012next}, including those shaped by lived experiences~\cite{LeDantec2009value}. Given its significant influence on technology design~\cite{friedman2013value,borning2012next,winkler2021twenty}, recent efforts have adopted this approach in the design of AI systems by identifying user values and tailoring AI systems to support diverse values beyond traditional metrics such as performance and usability, under the name of \textit{value-sensitive AI}~\cite{zhu2018value}. 

We position our research at the intersection of techno-spiritual traditions and human-AI interaction, incorporating VSD principles to advance both domains. For techno-spiritual design, we explore the potential of emerging AI technologies, such as large language models (LLMs), as a foundation for developing spiritual technologies. For value-sensitive AI and human-AI interaction, we investigate how integrating spiritual values can re-purpose AI, fostering more nuanced, human-centered designs for future AI systems.

In conducting our research, we focused on \textit{praying} because it constitutes an observable \textit{act}, one that reifies the abstract values that shape spiritual life. Praying is defined as an intentional act that is oriented toward transcendent beings, with the expectation of reception (or, being heard) \cite{cerulo2008name}. Praying, simultaneously, is a complex spiritual practice intertwined with multiple values that extend beyond traditional system design goals, such as efficiency or productivity. Existing spirituality literature outlines several praying-related values, such as confession or thanksgiving ~\cite{laird2004measuring, whittington2010prayer, poloma1991effects}. To expand the ongoing design and HCI discussion on prayer technologies~\cite{gaver2010prayer,wyche2008sun,claisse2024designing,smith2020cannot,smith2023thoughts,o2020community,o2022community}, our study identifies key values of prayer for Korean Christian believers and specific ways in which those values are enacted, further providing design implications for spiritual technologies. 

The core aim of our study is not to advocate for the introduction of AI systems into spiritual experiences as a design \textit{``solution''} to a user’s \textit{``problem''}; rather, we design and propose conceptual spiritual AI systems as a means to provoke reflection, reinterpretation, and reframing—both in thinking about techno-spiritual technologies through the lens of AI, and in rethinking human–AI interaction and relationships through the lens of spirituality, deeply resonating with the spirit of Research through Design (RtD) \cite{zimmerman2014research}. Accordingly, through the design and evaluation of four spiritual AI systems, we aim to critically examine the implications of introducing AI into the realm of spirituality, as well as to reimagine the relationships between humans and AI that emerge from such experiences.

To do so, we first conducted a diary and interview study to identify the values that shape meaningful prayer experiences, along with users’ practices and preferences in enacting those values. Building on these practices and preferences, we curated a design workbook—resembling speculative design scenarios \cite{auger2013speculative}—that featured four conceptual spiritual AI systems. We then presented these design concepts to users as prompts for co-exploration and reflection on the implications of introducing AI systems into spiritual contexts.

Our findings reveal that what participants value most in spiritual technologies is not instrumental support, but the preservation of \textit{authenticity} in their spiritual practice, often tied to subjective feelings of connection to a divine being. From this perspective, we critically interrogate how the presence of AI may alter such authentic experiences, specifically attending to how it might increase or diminish these feelings of \textit{authenticity}. Based on these findings, we further examine the role of AI in spiritual (and, by extension, other deeply personal and value-oriented) experiences. In doing so, we discuss how designers might preserve user agency and authenticity by leveraging AI’s inexplicability as a resource for personal meaning-making, while also acknowledging that there are realms in which AI’s presence may fundamentally undermine the core value of an experience, resonating with notions of AI non-use.

Our contributions are threefold:

\begin{enumerate}
    \item We empirically explore how values related to prayer are enacted and experienced by Korean Christian believers.
    \item By designing conceptual spiritual AI systems grounded in these values and practices, we examine how AI may enable or constrain prayer experiences.
    \item By engaging participants with these designed concepts, we reframe AI's goals beyond efficiency or accuracy, articulating how AI systems might instead recognize and preserve user agency.
\end{enumerate}

\section{Backgrounds}
\subsection{Techno-spirituality and Praying}

Techno-spirituality is a field that examines how one’s spiritual, transcendent, and religious experiences are or could be facilitated through technology~\cite{buie2019let}. The most prominent studies of techno-spirituality have been on online religious communities, where individuals gather to share their beliefs and often pray for one another~\cite{smith2020cannot, helland2007diaspora, smith2023thoughts, o2020community, o2022community, claisse2023keeping}. Other studies have focused on facilitating specific activities, such as connecting religious communities~\cite{hlubinka2002altarnation, wyche2006technology}, tangible artifacts for religious and spiritual purposes \cite{markum2023designing, markum2024mediating}, supporting religious activities in the forms of prayer counting \cite{claisse2024designing} or providing users with the necessary information to perform religious rituals such as possible prayer topics or reminder for prayers~\cite{woodruff2007sabbath, wyche2008sun, gaver2010prayer, biddlecombe2004cell}. 

In examining these concepts, many design and HCI scholars have sought to move beyond superficial task support or the mere replication of offline spiritual experiences, attending to how technology could more meaningfully support spiritual behaviors \cite{wolf2023god, wyche2009extraordinary}. By doing so, the techno-spiritual research community has not only expanded the scope of spiritual technology, but also shown the possibility of re-purposing technology’s traditional goals, such as efficiency, speed, or accuracy \cite{gaver2010prayer, bell2006no}. In line with this spirit, one thread in contemporary techno-spiritual research focuses on the deeply value-oriented nature of spiritual practices. Most notably, \citeauthor{wolf2024still} called for greater exploration of real-world applications of religious and spiritual technologies that more closely align with values inherent to spiritual activities, such as peace, sincerity, happiness, and connection \cite{wolf2024still}.

Such an emphasis is particularly significant in the context of praying, as many believers emphasize the inherently personal and reflective act that involves a profound connection between the individual and their spiritual values~\cite{whittington2010prayer, james2003varieties}. As a critical aspect of many people’s spiritual lives, the act of praying represents and intertwines with spiritual values such as comfort, guidance, connection with the divine, a sense of peace, and rituals within a religious belief system~\cite{laird2004measuring, whittington2010prayer, poloma1991effects, gaver2010prayer}. Reflecting these discussions, recent techno-spiritual study highlights prayer as a key aspect of first-person spiritual experience \cite{song2025walking}.

Our research builds on this rich tradition of techno-spiritual scholarship to identify the key values that shape the prayer experiences of Christian believers in South Korea, as well as the ways in which these values are enacted in practice. Building on this exploration, we examine how AI may support, reshape, or hinder such spiritual experiences.

\subsection{Values, Value Sensitive Design, and Value Sensitive AI}
\subsubsection{Values and Value Sensitive Design}
Values refer to “what a person or a group of people consider important in life” \cite{friedman2013value}, ranging from concrete everyday factors such as friends to abstract concepts such as virtue. Prominent examples of such values discussed in design and HCI include privacy \cite{wong2017eliciting}, autonomy \cite{friedman1996user}, or trust \cite{vermaas2010designing}.

Resonating with this strong tradition of incorporating value dimensions in design and HCI, \citeauthor{friedman2013value} introduced Value-Sensitive Design (VSD), a systematic approach to technology design that prioritizes human values~\cite{friedman2013value}. VSD integrates conceptual, empirical, and technical methods to create technologies that reflect the values of users and broader stakeholders. The process begins with a conceptual investigation, where researchers identify specific stakeholders, their values, and potential value conflicts. These insights are then empirically validated by gathering data on users' value-oriented experiences with technology. Finally, technical investigations explore how systems can be designed to address the identified user values while adhering to the proposed design constraints \cite{boyd2022designing}.

The strong theoretical grounding of this design approach has influenced many researchers who incorporate values into their design work~\cite{foong2024designing, wong2023broadening, yoo2013value}. Since the suggestion of VSD in 1996, there have been more than 200 papers using VSD, as reviewed by Winkler and Spiekermann~\cite{winkler2021twenty}. Notably, \citeauthor{LeDantec2009value} argued that values are situated in contexts, highlighting the limitations of conceptual, top-down approaches in early VSD work~\cite{LeDantec2009value}. They emphasized the importance of reflecting on local contexts to understand the values that actual users care about in their daily lives. \citeauthor{borning2012next} took a step further in suggesting the need to tone down universal values to better account for the situatedness of values~\cite{borning2012next}.

\subsubsection{Value Sensitive AI}
With recent advancements in AI, scholars have increasingly emphasized the importance of foregrounding human values in the design of AI algorithms and AI-assisted technologies~\cite{sadek2024guidelines,sadek2024value}. As \citeauthor{umbrello2021mapping} argue, because we lack a clear understanding of how AI systems acquire knowledge through machine learning mechanisms, it becomes crucial to attend to how values are learned and embedded within these systems, underscoring the need for novel value-sensitive approaches in the field of human-AI interaction~\cite{umbrello2021mapping}.

Notably, \citeauthor{showkat2023right}'s literature review on algorithmic homelessness service provision research revealed that, even in studies aimed at supporting marginalized individuals, values related to academic productivity, such as novelty and performance, dominate, while more urgent values such as inefficiency, violated privacy, and the reproducibility of homeless conditions are often overlooked~\cite{showkat2023right}. Based on these critical reflections on AI research, emerging scholarship has proposed approaches to account for stakeholder and community values. \citeauthor{zhu2018value} proposed Value-Sensitive Algorithm Design, where community stakeholders’ values are incorporated into the machine learning pipeline, allowing stakeholders to guide critical decisions throughout the process~\cite{zhu2018value}. This approach also highlights the iterative nature of value-sensitive design for AI. Stakeholder involvement and iteration have been suggested in a value-based framework for the evaluation of AI systems~\cite{yurrita2022towards}, as well as the consideration of context-specific values \cite{liscio2021axies}.

AI alignment is another key topic in discussions around AI and human values, referring to how AI systems should operate in alignment with users' values, ensuring they are helpful, safe, and reliable~\cite{terry2023ai}. Although its roots lie in philosophical approaches to AI systems~\cite{gabriel2020artificial}, AI alignment has gained substantial attention in machine learning and AI research~\cite{shen2024towards,terry2023ai,suh2024luminate}. AI alignment often concerns how AI systems align with human values either during development (e.g., through annotation) or interaction (e.g., via human feedback)~\cite{li2023coannotating,chung2022talebrush}. \citeauthor{shen2024towards} proposed the notion of bidirectional alignment, which includes HCI efforts in AI education, fostering critical thinking about AI, and human-AI collaboration~\cite{shen2024towards}. 

We aim to contribute to this growing body of value-sensitive AI and AI alignment research by examining how AI may both aid and hinder the deeply value-laden act of praying. In particular, we foreground how AI’s presence and intervention can either support or undermine experiences of \textit{authentic} prayer, and how AI-enabled spiritual systems might preserve such authenticity by recognizing, respecting, and expanding users’ agency within human–AI relationships.

\subsection{Epistemology: Design Research and Research through Design}
Before proceeding to our main study, we situate this work within the rich tradition of design research. Unlike natural or social science approaches that often ground accountability in generalizability or reproducibility through emphasis on sample representativeness, scale, or methodological rigor to uncover a universal truth, design studies are often guided by aesthetic accountability of finding what ``works'', asking the question of what \textit{could be} rather than what \textit{is}. As \citeauthor{gaver2014science} notes, “Science uncovers what exists, and design creates the new.” Hence, we approach design as a means of exploring possibilities and expanding the scope of what is imaginable through making and reflection \cite{gaver2014science}.

\begin{figure*}[t]
\centering
\includegraphics[width=\linewidth]{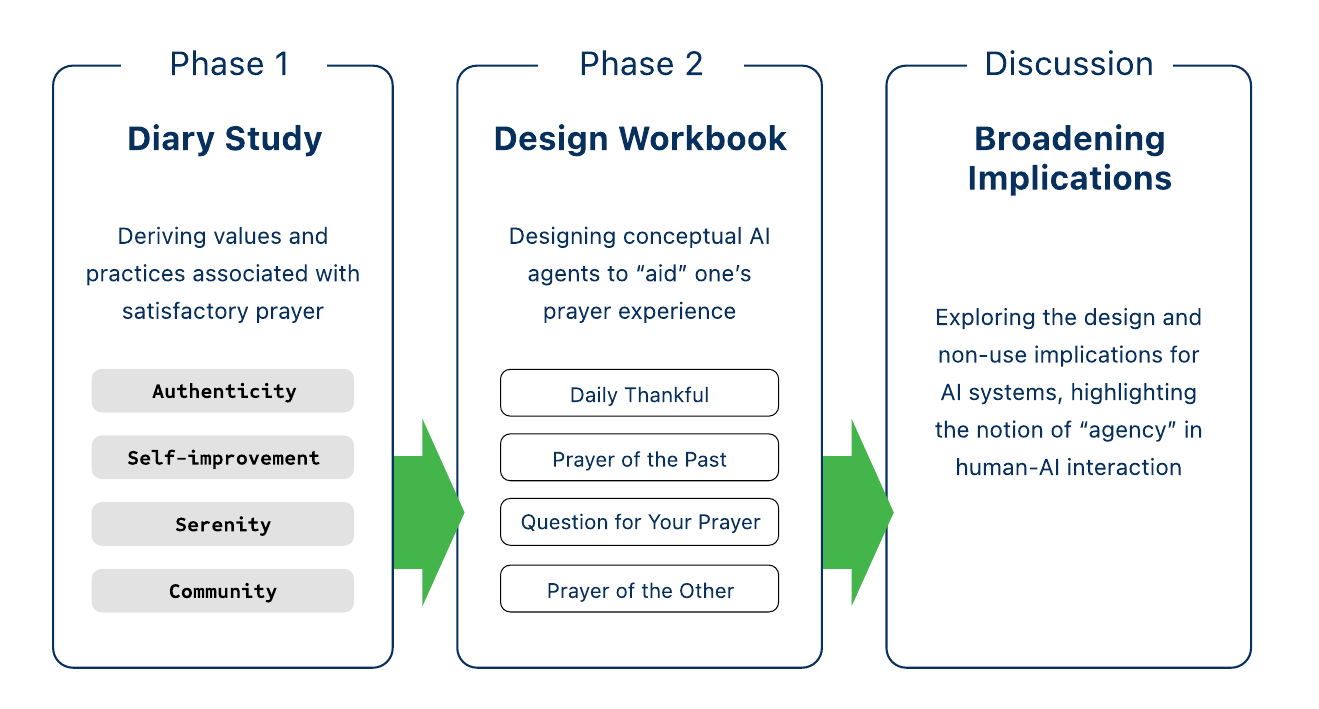}
\caption{Overview of Study Process}
\label{fig:overview}
\Description{The figure presents a pipeline connecting empirical findings to design and reflection. In Phase 1, a diary study identifies four values associated with meaningful prayer experiences: authenticity, self-improvement, serenity, and community. In Phase 2, these values inform the design of four conceptual AI systems. Daily Thankful, Prayer of the Past, Question for Your Prayer, and Prayer of the Other. In the final discussion stage, the study examines how these systems support or challenge the identified values and considers broader implications, particularly the role of agency and the possibility of non-use in human-AI interaction.}
\end{figure*}

This epistemological grounding is important to our work because we do not view the goal of design research as universalizing a particular experience made possible through technology. For instance, while one participant described manually entering prayers into the system as facilitating reflection, another found the same process tedious and burdensome. Even if a majority of participants favored one experience, that would not invalidate the other. Rather than treating divergent experiences as noise or exceptions, we see them as generative points of inquiry. Even when a design resonates with only a small number of people, it can still offer valuable insight into how and why certain designs take on meaning within particular populations, contexts, or situations. In this sense, our approach aligns with perspectives that understand design knowledge as accommodating multiplicity and plural experiences, rather than converging on a single authoritative account \cite{bardzell2010feminist}, and contrasts with paradigms in which truth is often treated as exclusive or singular \cite{kuhn1970structure}.

This epistemological grounding also resonates strongly with the ethos of Research through Design (RtD). RtD differs from design practices that emphasize solution-oriented outcomes, such as design research that seeks to derive prescriptive guidelines. Instead, RtD foregrounds the production of knowledge through the act of designing itself, using new artifacts to disrupt, complicate, or reframe existing understandings of the world \cite{zimmerman2014research}. Whereas research \textit{into} design examines design as a theoretical field, and research \textit{for} design focuses on producing design outcomes or prescriptive considerations, RtD is concerned with making new things to unsettle, complicate, or transform the current state of the world \cite{zimmerman2014research}. 

Accordingly, our aim is not to prescribe what “spiritual AI” should look like through universal design principles \cite{dourish2006implications}, or to create a concrete design artifact ready for use. Rather, we use conceptual designs of spiritual AI as a means to provoke reflection and discussion about how the introduction of AI into deeply value-oriented experiences might be experienced by people. Through this process, we critically examine both the possibilities and limits of AI in such contexts. We invite readers to engage with this work from this epistemological position, approaching the designs not as solutions to be evaluated for correctness, but as prompts for thinking with, questioning, and reimagining the role of AI in spiritual life.

\section{Overview}

In this section, we provide a brief overview of the study we conducted. In the first phase, eight Korean Christian participants took part in a two-week diary study, along with pre- and post-interviews, where they recorded their daily praying experiences, including what they prayed about and how they prayed. A reflexive thematic analysis identified four key values in prayer--authenticity, self-improvement, serenity, and community--along with the specific practices participants used to realize them.

In the second phase, we created four conceptual AI-based prayer assistance designs based on the identified values and practices. We distributed a design workbook introducing each concept and its use-case storyboard to 21 Korean Christian participants and conducted structured interviews, where participants shared their reactions and feedback on each concept. A reflexive thematic analysis yielded several themes that critically explore the implications of AI agents entering the realm of prayer, which is deeply entangled with nuanced human values that extend beyond those typically discussed in AI system development, such as efficiency or accuracy.

\textit{\textbf{Positionality Statement}: Before moving on to the main part of our study, we provide an author positionality statement to transparently communicate how our perspectives may have influenced the study planning, execution, and presentation. All authors except the third author are South Korean nationals. Of the four authors, the first and fourth identify as Christians, with the first Catholic and the fourth Protestant. The first author led most aspects of the study, including conceptualization, study design and execution, concept design, analysis, and writing. His experiences as a Korean Catholic--including early Catholic education, extensive periods of non-practice, engagement with Protestant values through undergraduate coursework at a Protestant university, and later baptism during military service around the time of executing this research--have both informed and limited the scope and presentation of this study. Specifically, he acknowledges having relatively limited lived understanding of communal practices within Korean Christian communities beyond his experiences in military Catholic churches and communicates these limitations in the analysis and presentation of this work.}

\section{Phase 1: Diary and Interview Study}
\subsection{Diary Study}
The first phase of our study aimed to elicit key \textit{values} related to a satisfactory praying experience, as well as users' \textit{practices} in realizing them. These values and practices were drawn with the goal of orienting the conceptual spiritual AI systems for the design workbook. To achieve this, we conducted a 2-week diary study with 8 participants, accompanied by pre- and post-interview sessions. 

A diary study is a user research method where participants self-document their behaviors, thoughts, and experiences during specific occasions. Our choice of the method was informed on how: (1) the entries itself would provide insights into the users' experiences and values surrounding their praying experiences, (2) the act of writing would prepare users for the interview by prompting reflection on their experiences~\cite{chen2010diary, sohn2008diary, palen2002voice, carter2005participants, jokela2015diary}, (3) the method has proven successful in many formative, exploratory studies~\cite{rieman1993diary, colbert2001diary, hayashi2011diary, chen2010diary}, and (4) it is particularly useful for investigating private, non-observable situations, as they are conducted by the users within their situated environments~\cite{czerwinski2004diary, chen2010diary, colbert2001diary}.

\subsubsection{Recruitment}
To explore values related to a fulfilling prayer experience, we initially refrained from restricting recruitment based on specific religious affiliations, racial backgrounds, genders, or age groups. To ensure representation from smaller religious denominations in South Korea, where the study was conducted, we distributed recruitment materials in both Korean and English across various religious communities within the university. Participants were required to meet two criteria: (1) engage in prayer at least five times per week and (2) self-report that their faith plays a significant role in their lives, as defined by prior research on praying experiences~\cite{krause2000using}.

Despite these efforts, all recruited participants were Christians aged 20 to 28, with one identifying as Catholic and the remaining seven as Protestant. We speculate that such a demographic outcome reflects the religious landscape of South Korea, where Buddhism and Christianity dominate the religious population, with virtually no Muslim representation. Furthermore, we speculate that Christian individuals tend to align more closely with our inclusion criteria of praying regularly, as Christian beliefs often place a greater emphasis on prayer as a spiritual obligation. Consequently, we present our findings within the situated experiences of Korean Christians, seeking transferable insights for design rather than representative findings or qualitative saturation~\cite{haraway2013situated, gaver2014science, soden2024,hayes2011relationship}.

\subsubsection{Study Process}
We initiated our study with individual in-person introductory interviews. These semi-structured interviews, each lasting approximately an hour, had three objectives: (1) to acquaint participants with the project's purpose and their role, (2) to establish an understanding of their values and practices related to praying, (3) to ask their preferred media for documenting their experiences for the diary entries (e.g., handwriting, typing, voice recordings, etc.) before initiating the diary study, and (4) to build rapport between participants and researchers~\cite{rieman1993diary}. All sessions were recorded with the participant's consent. At the end of each session, we reaffirmed their willingness to participate in the diary study and provided them with a detailed study timeline. All eight participants graciously agreed to participate in our research endeavor. Detailed personal information for the participants is presented in Table \ref{tab:participants}. 

\begin{table}
  \caption{Participant Details for Diary Study}
  \Description[table]{Eight participants (P1–P8) in the diary study, mostly Protestant with one Catholic, ages ranging from 20 to 28. Seven participants are female, and one is male. The number of diary entries per participant ranges from 10 to 26.} 
  \sffamily
  \small
  \label{tab:participants}
  \centering
  \renewcommand{\arraystretch}{1.1}
  \setlength{\tabcolsep}{4pt}
  \rowcolors{2}{gray!15}{white}

  \begin{tabular}{
    >{\centering\arraybackslash}p{0.08\linewidth}
    >{\centering\arraybackslash}p{0.18\linewidth}
    >{\centering\arraybackslash}p{0.10\linewidth}
    >{\centering\arraybackslash}p{0.14\linewidth}
    >{\centering\arraybackslash}p{0.20\linewidth}
  } 
    \toprule
     & \textbf{Religion} & \textbf{Age} & \textbf{Gender} & \textbf{Diary Entries}\\
     \midrule
    \rowcolor{white}
    P1 & Protestant & 28 & Female & 26 \\
    P2 & Protestant & 20 & Female & 12 \\
    P3 & Catholic   & 25 & Female & 15 \\
    P4 & Protestant & 23 & Female & 10 \\
    P5 & Protestant & 24 & Female & 19 \\
    P6 & Protestant & 26 & Female & 11 \\
    P7 & Protestant & 24 & Male   & 10 \\
    P8 & Protestant & 22 & Female & 10 \\
    \bottomrule
  \end{tabular}
\end{table}

For two weeks, participants were instructed to compose at least five diary entries per week. We asked them to write an entry whenever they engaged in prayer or experienced moments they deemed divine. Multiple avenues were provided for participants to share their entries, but all chose either a shared diary application or a mobile messenger platform to write their entries digitally. Each entry was to include the following: (1) the context of the prayer (e.g., commuting on the bus to work), (2) the intended recipient of the prayer (e.g. God), (3) artifacts used, (4) technological tools or services used for the prayer, (5) the content of the prayer itself, (6) their emotions before, during, and after the prayer, (7) what an imaginary magical tool would do to enhance their prayer experience, and (8) any supplementary remarks they wished to share. 

Our prompts were deliberately designed to include both open-ended sections that allowed participants to freely share the experiences they deemed important, as well as closed questions to identify specific values and practices that we were looking for~\cite{colbert2001diary, sohn2008diary}. We emphasized that participants should write their entries as soon as possible after the event to ensure immediacy and accuracy~\cite{jokela2015diary}. If they later realized something additional about an experience, they were encouraged to edit or add to the entry, indicating that the notes were written afterward. To prevent unnecessary exploitation, we emphasized multiple times that they could omit private parts of their prayers and only asked them to share to the extent they deemed appropriate. 

In preparation for the final interview, we devised personalized questions for each participant based on their written diary entries. For instance, for a diary entry that said “the fire that was ignited in me last week still has not extinguished, so I inquired why this was. I was responded with the above image,” (P1) we asked her to further elaborate on how she conceptualizes such “response” from the divine, as well as how she encounters and interprets them. This approach aimed to ensure a deeper qualitative and contextual understanding, as relying solely on written entries can sometimes yield superficial insights~\cite{gaver1999design, crabtree2003designing}. Conducted as individual semi-structured sessions, the final interviews lasted approximately 60 minutes each. The objectives of these interviews were: (1) to uncover diverse values and specific practices users utilize to support those values during prayer, and (2) to explore potential avenues where AI could support such values and practices. All interviews were recorded with the participants' consent.

Upon completing the introductory interview and the first week of the diary study, participants were compensated with 15,000 KRW (11.20 USD); after the second week of the diary study and the final interview, they received an additional 35,000 KRW (26.20 USD). The entire study process was approved by the institution’s ethics review board.

\subsubsection{Analysis}
Drawing from the collected diary entries and transcripts of both the introductory and final interviews, we engaged in a reflexive thematic analysis (RTA) \cite{Braun2021thematic}. RTA is a qualitative analysis method that foregrounds researchers’ agency in interpreting data and the messy, iterative nature of analysis, rather than relying on rigid analytic structures.

Following its steps, the first author read through both forms of data multiple times to become familiar with the material. He then inductively coded the data using both inductive, semantic codes that captured surface-level observations (e.g., when and where participants prayed) and deductive, latent codes that noted implicit interpretations (e.g., how deep reflection leads to a satisfying prayer experience). The first and second authors then collaboratively developed themes from these codes through multiple iterations. These themes included participants’ definitions of what constitutes a good prayer, different types of prayers, prayer practices, media used during prayer, and factors that foster or hinder a satisfactory prayer experience.

Based on these themes, the first and second authors further discussed the underlying values reflected in these practices and conditions. In other words, we examined how certain behaviors (e.g., keeping a prayer journal) or preferences (e.g., favoring a secluded, private environment over an open space) point to particular values. For instance, the two examples led to the value of \textit{self-reflection}, where participants recount daily events and engage in honest, quiet reflection without the presence of others. These procedures ensured that our identified values were grounded in participants' situated realities, resonating with \citeauthor{LeDantec2009value}’s VSD approaches \cite{LeDantec2009value}.

\subsection{Findings: Prayer Values and Practices}
Through our analysis, we identified four values related to praying experiences: self-reflection, serenity, community, and authenticity, as well as specific user practices to realize them. Below, we present our findings, especially highlighting the grounds that guided us in creating the design concepts in the second part of our study. 

\subsubsection{\textit{Self-Reflection}: A Good Prayer is a Process of Becoming a Better Person}

One of the most prominent themes that emerged was the strong connection between prayer and self-reflection. For our participants, praying served as a moment to revisit their daily experiences and recount them to the divine, prompting introspection and helping them identify areas for self-reflection. Participants reported several practices they used to pursue this self-reflection. For instance, P2’s diary entry showcases how she reflected on her day to improve herself as a teacher:

\begin{quote}

\textit{``I go to lessons still lacking in many ways. Please guide me to recognize my shortcomings as I teach and to become someone who continually works to improve them.''} (P2, Diary entry 15)

\end{quote}

One notable practice involved treating prayer as a ``pact’’ between themselves and the divine. For instance, a prayer might include a promise to refrain from certain behaviors. These pacts motivated participants to follow through on their commitments, offering a stronger sense of accountability than vague resolutions such as “I should be more diligent'' or “I should help those in need.''

In line with the value of self-reflection, keeping a list of prayer topics to remember what to pray for was another common practice among participants. Throughout the day, they noted topics to bring into prayer, particularly those related to daily regrets and moments of gratitude. During prayer, these lists served as prompts for reflection, and participants often revisited past entries to track their progress over time.

Participants also emphasized the importance of confronting discomfort during prayer. Many expressed frustration with self-reflective or confessional prayer, which they associated with feelings of guilt and shame. These emotions sometimes led them to avoid praying altogether or to rely on reciting pre-written texts. Yet, by working through this discomfort, participants described experiencing meaningful self-reflection and a renewed motivation to act on it.

Altogether, the core of self-reflection exists in participants taking active agency over examining their daily lives. Rather than being told how to live, they used prayer as a tool to encourage self-reflection through various practices. The importance of agency in self-reflection was highlighted by P7’s remark that overly passive forms of prayer can hinder personal growth, resulting in an unsatisfying prayer experience. After learning about our study goals, P7 became curious about using AI for prayer. This curiosity led him to try a ChatGPT-based prayer assistance system, which generated fully written prayer texts based on his prompts:

\begin{quote}
\textit{``When I used ChatGPT, it wasn't satisfactory. It wrote everything about my situation for me. Prayer is an experience, and we need time to process that while reciting. But when I used the Bible application, it asked questions like 'What do you think about this?' 'What experiences do you have on this matter?' I liked it because it helped me and prompted me to brainstorm about what to think about certain Bible verses.'' }(P7, Post-Interview)
\end{quote}

P7 found that AI-generated prayer texts removed the space for personal contemplation, thus leading to a prayer where he was unable to take agency in self-reflection. In contrast, P7 was more content using a Bible application he had used previously, which posed questions about specific Bible verses, prompting deeper reflection. Despite being more demanding, this method offered a deeper, more satisfying prayer experience.

In summary, participants highlighted various practices that underscore the value of self-reflection through prayer. In doing so, they particularly emphasized the importance of taking agency, actively engaging in self-reflection rather than being rendered passive to pre-written prayers or those written for them.

\subsubsection{\textit{Serenity}: Reaching a Tranquil Emotional State}
Another important value was achieving emotional serenity through prayer. Participants often found themselves praying in challenging situations, either directly requesting the divine to grant their desires or seeking guidance from the divine toward the most suitable course of action. In both cases, participants emphasized that the primary objective was not to have their wishes granted; the most significant satisfaction came from the relief of disclosing their concerns and frustrations to a powerful being and unburdening themselves. As P2 shared: 

\begin{quote}
\textit{``Even if it's an echo to the void without an answer, you get a sense of relief just by saying it to a certain figure [...] The feeling of relief would be the biggest purpose (of prayer).'' }(P2, Introductory Interview)
\end{quote}

\begin{quote}

\textit{``I feel relieved being able to easily open up about concerns I cannot share with others…''} (P2, Diary entry 7)

\end{quote}

Even when those concerns did not directly go away, they found consolation in the fact that the divine knows their situations. The idea that the problem, once they shared it with the divine, was part of a bigger plan ultimately led to less emotional distress.

Further, participants found serenity by interpreting ``answers'' to their prayers in ordinary daily events. Many recalled ``receiving answers'' through remarks their friends and family made in conversation, phrases from books they were reading, or other serendipitous encounters. For instance, P6 shared the following diary entry:

\begin{quote}

\textit{``I spent a lot of time thinking about how I should continue to hold on to a prayer request that had not been answered (…) Over the past week, I had been praying with this concern in mind, and I felt that a Bible verse quoted in the sermon came to me as an answer, so I offered a prayer of gratitude. Although it did not seem entirely clear and I wondered whether I had interpreted it correctly, I was still thankful, as it felt like a message I needed.''} (P6, Diary entry 7)

\end{quote}

Had she not prayed, these moments would have simply passed by. However, the prayer became a lens through which they observed the world, allowing them to actively interpret their everyday experiences, discover unforeseen solutions to issues they had articulated in prayer, and attain a sense of peace.

In short, in the pursuit of serenity, participants used prayers not only to unburden their concerns and supplicate to a higher power, but also to interpret and recognize ``answers`` to their prayers in everyday interactions, helping them find solace.

\subsubsection{\textit{Community}: A Sense of Connection with the Religious Community. }
Another important value participants discussed was the sense of community with those who share the same belief. Unlike the other values, which were largely seen as desirable, community carried ambivalent meanings, as participants expressed contrasting views on whether sharing one’s prayer experiences felt fulfilling or not.

On the positive side, they experienced both functional and emotional enhancements through communal experience. Functionally, participants also described moments of praying together, where the awareness of others observing their prayers encouraged greater diligence. These shared practices took various forms, both offline and online. Notable uses of digital sharing include Zoom prayer sessions during COVID-19, where people took turns praying or prayed simultaneously, and online chatrooms where participants shared their prayer journals every day. Participants reported that these interactions fostered a ``pleasantly forced'' state, prompting them to commit to their praying activities. Furthermore, listening to other members' prayers allowed them to explore novel prayer methods or topics to pray for, expanding the scope of their prayer experience.

Emotionally, sharing prayers deepened solidarity within religious communities, fostering a sense of belonging through resonance and unity in collective experiences~\cite{smith2021what, kaur2021sway}. P1, in particular, shared how an embodied experience of praying together became a powerful source of their religious solidarity: \textit{``If a community prays together out loud, looking at each other's faces, breathing together, listening to what others are praying, [...] sometimes praying for each other, listening to that heals and motivates, realizing that we all are together.''} (P1, Final Interview)

Conversely, some participants hesitated to share their prayer experiences with others, finding that the presence of an audience hindered authentic prayers and deep self-reflection. P4 and P7 notably shared:

\begin{quote}
\textit{``The (Korean) Christian community has a strong judgment on what is a good prayer. It can't be too short, you have to use the right word in the right way (...) there are these standards. I think, regardless of such format, what is important is communicating with the divine, but I hate to do it in front of other people in the church because I know that it's going to be judged.'' }(P7, Pre-Interview)
\end{quote}
\begin{quote}
\textit{``Sometimes when I pray in church, I cry. But when someone sees that, the moment I realize that, my prayer and mind start to shake. It's nothing to be ashamed about, and I know that, but I just do (...) The possibility of someone being able to see me hinders me from going deep into the prayer. As the prayer gets deeper, you need to bring out your raw mind, but I just can't do that.''} (P4, Post-Interview)
\end{quote}

As illustrated above, awareness of the audience sometimes led participants to feel less immersed in their prayers or to include content intended to appease listeners rather than to express their authentic sentiments. This was especially common in conservative religious communities, where adhering to particular formats or expressions is considered essential for prayer.

In short, a sense of connection with others who shared the same beliefs was a key value in prayer, operating both positively and negatively. Praying together enhanced both functional and emotional dimensions: functionally, it increased diligence and introduced participants to new methods and topics; emotionally, it deepened solidarity and fostered a sense of belonging. However, some participants felt that the presence of an audience hindered sincerity and limited deeper self-reflection, particularly in more conservative settings.

\subsubsection{\textit{Authenticity}: Preserving a Genuine Connection with Divine. }
Perhaps most importantly, participants valued an \textit{authentic} experience while praying. They described authenticity primarily as sincerity and honesty in what they expressed during prayer and, more broadly, as a deep, natural connection with the divine and honesty toward it.

Of particular relevance to HCI audiences, participants explicitly discussed how this value could be disrupted or diminished by technological intervention. Many expressed discomfort with the idea of technology mediating their spiritual experiences, noting that it felt like introducing an artificial barrier or detracting from the personal and intimate nature of their relationship with the divine.

As Christians, many of our participants viewed prayer as a means to foster a closer connection with the divine, describing prayer as a ``conversation'' with God. They drew parallels with writing a letter or an email, where the response might be delayed or absent, yet the act of composing assumes a recipient and can be seen as a form of dialogue. In this sense, many actively sought to feel the presence of divine power, often visualizing how their prayer would be heard by God or how the Holy Spirit would physically interact with them to draw prayer from them.

Within this context, many participants found technologies that mediate their connection to the divine to be unnatural, unnecessary, and uncomfortable. For instance, P4 expressed that:

\begin{quote}

\textit{``to think that technological or scientific intervention is taking place in the realm of prayer, it feels like damaging a part of nature or hindering the moment where I return to that original, natural state'' }(P4, Post-Interview). 

\end{quote}

P4 further described how they found such technological aid unnecessary, given how well-established their relationship with the divine already is:

\begin{quote}
\textit{``I once saw a worship where people used lasers and drones to project the form of Jesus in the sky. I was very uncomfortable, even though it's nothing to be bothered about, since there are many instances where they use the image of Jesus. I think the reason is that it felt unnatural to me. Even without technological intervention, my spirituality and mind could still rise and burn, but that felt like pouring technology to artificially make me burn even more.'' }(P4, Post-Interview)
\end{quote}

Participants were also reluctant to use technology because they did not believe it could reenact or provide the ambiguous elements that allow them to immerse themselves in prayer. For instance, when discussing spatial elements that contribute to satisfactory prayers—whether in a darkly lit room, a grand church, in the middle of nature, or a crowded subway—they expressed skepticism about current technology's ability to recreate these environments. They further voiced that, even if a system could replicate the scent, temperature, humidity, scenery, and luminosity, the very idea that they are not \textit{actually} present would hinder their satisfaction. P3 mentioned that she ``would not entirely immerse myself, thinking that 'this is done by an AI' in a small part of (her) mind.'' (P3, Post-Interview)

Nevertheless, participants also acknowledged the contemporary reality where many religious individuals and authorities continue to leverage technology for religious purposes, such as using mobile Bible applications, PowerPoint presentations during services, participating in video-call prayer groups, or using AI for prayer support. These systems allowed them to overcome certain constraints (such as physical locations) to participate in religious activities or to avoid tedious tasks (such as flipping through scriptures to find a verse). P7 discussed how ``before the pandemic, holding remote worship was considered heresy'' (P7, Post-Interview) and anticipated that further development of human-technology relationships might help us understand what technology should and should not do in the realm of spirituality.

In short, while most participants expressed reluctance to use technologies to aid or augment their deeply spiritual prayer experiences—finding them unnatural, unnecessary, or uncomfortable—they nevertheless described various practices of using technology in different contexts. Connecting this observation to the value of authenticity, participants appeared to continually negotiate the degree of technological presence that would not hinder their authentic experience while still allowing them to leverage the benefits of technology within a rapidly changing landscape of religious and spiritual tools.

\section{Phase 2: Design Workbook Study}
Based on the values and practices identified in the first phase of our study, we designed four conceptual AI systems intended to “aid’’ praying experiences. We compiled these designs into a design workbook and discussed the concepts with participants. 

A design workbook refers to a compilation of design concepts and proposals, often used to provoke viewers to discuss, reflect, and speculate on the values, goals, and potential usages of the given design~\cite{gaver2011making, wong2017eliciting}. Having its roots in speculative design \cite{dunne2013speculative}, the goal is not to create an actual solution for immediate use but to stimulate discussion about the context of use, values, and ideas related to a certain form of technology~\cite{tanenbaum2012steampunk, blythe2014research, dunne2013speculative, linehan2014alternate, bleecker2015design, lindley2015back}. 

\begin{figure*}[h]
\centering
\includegraphics[width=\linewidth]{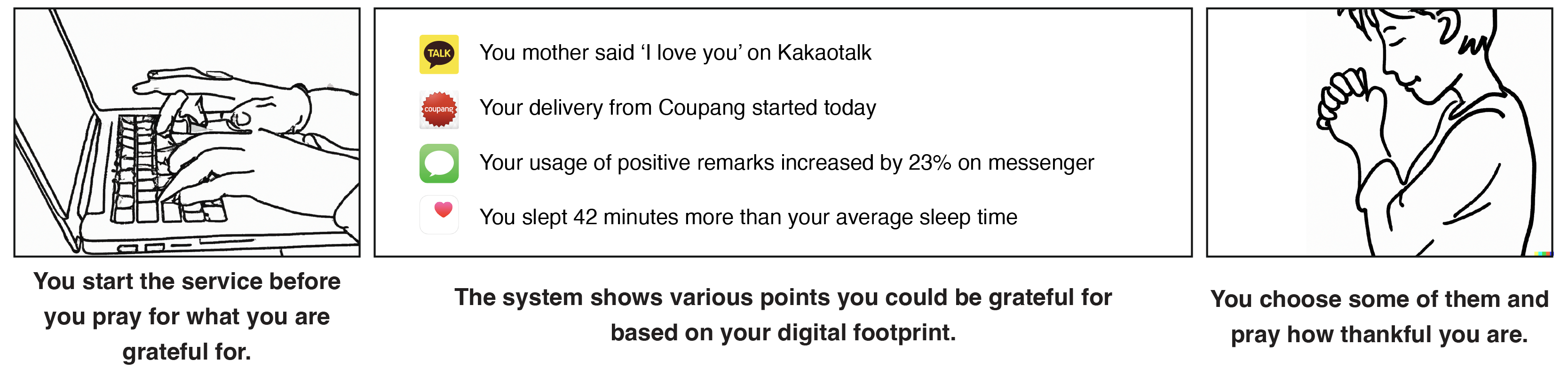}
\caption{Use Case Storyboard for \textit{Daily Thankful}}
\label{fig:dailythankful}
\Description{You start the service before you pray for what you are grateful for. The system shows various points you could be grateful for based on your digital footprint. The examples are as follows: ``Your mother said `I love you' on Kakaotalk. Your delivery from Coupang started today. Your usage of positive remarks increased by 23 percent on Messenger. You slept 42 minutes more than your average sleep time.'' You choose some of them and pray how thankful you are.}
\end{figure*}

\begin{figure*}[h]
\centering
\includegraphics[width=\linewidth]{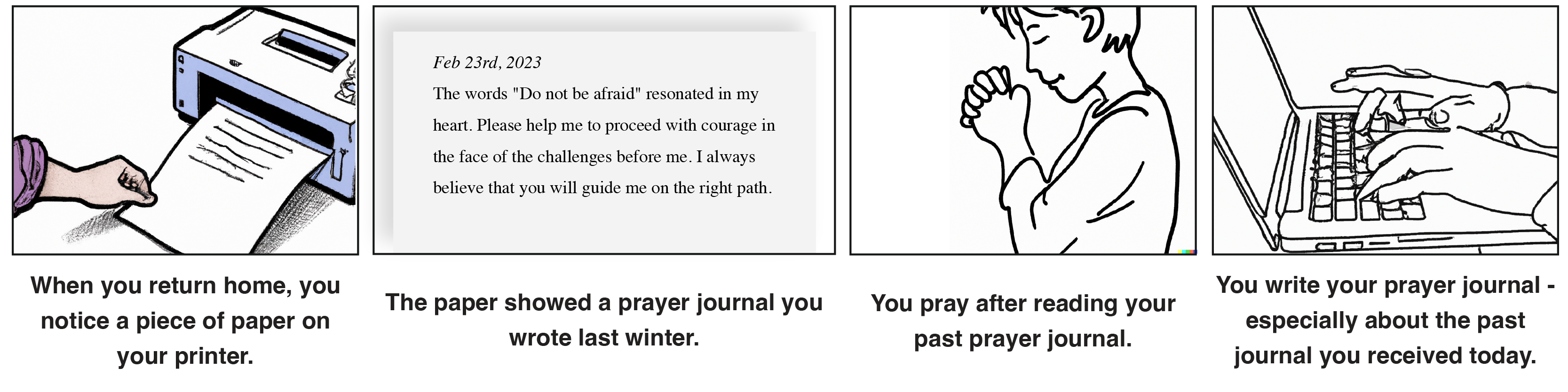}
\caption{Use Case Storyboard for \textit{Prayer of the Past}}
\label{fig:prayerofthepast}
\Description{When you return home, you notice a piece of paper on your printer. The paper shows a prayer journal you wrote last winter. It says ``Feb 23rd, 2023. The words ``Do not be afraid'' resonated in my heart. Please help me to proceed with courage in the face of the challenges before me. I always believe that you will guide me on the right path.'' You pray after reading your past prayer journal. You then write your prayer journal - especially about the past journal you received today.}
\end{figure*}

\begin{figure*}[h]
\centering
\includegraphics[width=\linewidth]{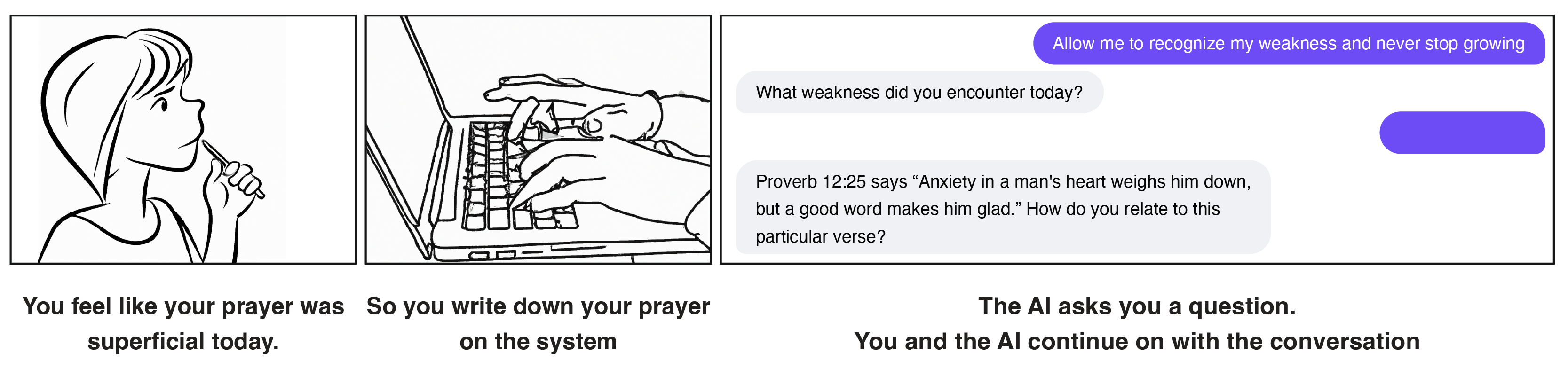}
\caption{Use Case Storyboard for \textit{Question for Your Prayer}}
\label{fig:questionforyourprayer}
\Description{You feel like your prayer was superficial today. So you write down your prayer on the system. The AI asks you a question. You and the AI continue with the conversation. The conversation goes like this: Allow me to recognize my weakness and never stop growing. What weakness did you encounter today? Proverb 12:25 says, “Anxiety in a man's heart weighs him down, but a good word makes him glad.” How do you relate to this particular verse?}
\end{figure*}

\begin{figure*}[h]
\centering
\includegraphics[width=\linewidth]{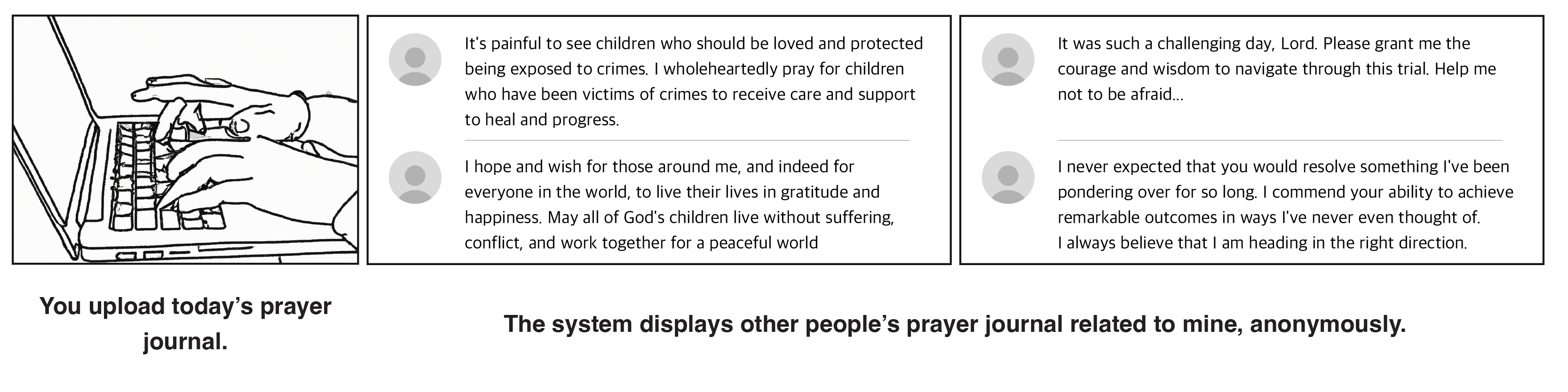}
\caption{Use Case Storyboard for \textit{Prayer of the Other}}
\label{fig:prayeroftheother}
\Description{You upload today’s prayer journal. The system displays other people’s prayer journals related to mine, anonymously. The example prayers are as follows: ``It's painful to see children who should be loved and protected being exposed to crimes. I wholeheartedly pray for children who have been victims of crimes to receive care and support to heal and progress.'' ``I hope and wish for those around me, and indeed for everyone in the world, to live their lives in gratitude and happiness. May all of God's children live without suffering and conflict, and work together for a peaceful world'' ``It was such a challenging day, Lord. Please grant me the courage and wisdom to navigate through this trial. Help me not to be afraid...'' ``I never expected that you would resolve something I've been pondering over for so long. I commend your ability to achieve remarkable outcomes in ways I've never even thought of. I always believe that I am heading in the right direction.''}
\end{figure*}

In our study, we created a 4-page design workbook consisting of a concise one-page pamphlet for each conceptual service. Each pamphlet featured the design's title, a brief introduction, a use-case storyboard (See Figures ~\ref{fig:dailythankful}, ~\ref{fig:prayerofthepast}, ~\ref{fig:questionforyourprayer}, and ~\ref{fig:prayeroftheother}), and overarching purposes behind each design. We disclose that some of the image materials for the storyboard were generated with DALL-E, a generative AI tool.

By presenting concrete design ideas rather than abstract notions, our goal in this phase was to explore how participants perceived different forms of AI presence in supporting prayer, ultimately offering insights into what AI’s role could be, and should not be, in this deeply value-laden experience.

In this section, we first detail the study process, including recruitment, the procedures for distributing the design workbook and gathering insights, and the analysis. We then present the four conceptual designs and the process for creating them.

\subsection{Design Workbook Study}

\subsubsection{Recruitment}

To conduct our study, we recruited 21 Korean Christian participants (11 female, 10 male; aged 22–39) from online communities. Of these, five had also participated in the earlier diary and interview study. Participants were individually invited to an online session, during which they were presented with the design workbook and asked to share their opinions. The decision to conduct the session online was made to increase convenience, and we determined that reading the workbook on screen would not significantly differ from reading a hard copy.

\subsubsection{Study Process}

We first presented the design workbook and asked participants to take as much time as needed to read it. Once they had finished, for each concept, we asked: (1) whether they were inclined to use the proposed system, (2) what they perceived as its positive and negative aspects, and (3) any additional comments they had. The order of presenting and discussing the concepts remained consistent across all sessions. Upon completion of the interviews, each participant received a gift card valued at 4,500 KRW (3.41 USD). Participant details and their inclination to use the proposed systems are shown in Table \ref{tab:participants_phase2}.

\subsubsection{Analysis}

In our analysis, the values from the previous phase functioned as an analytic lens for interpreting participants’ responses across the four systems. We adapted the original 6-step RTA \cite{Braun2021thematic} in four steps. First, the first author familiarized himself with the interview transcripts. Second, these responses were inductively coded based on the four values identified in the previous phase; for example, a comment about preferring to disclose one’s private matters to an AI rather than a priest, due to the absence of social judgment, was coded with ``authenticity.’’ Third, within each value category, all remarks associated with the same semantic code (i.e., value) were revisited to identify implicit meanings within them, where such interpretations were then assigned with deductive, latent codes. Finally, the first and second authors collaboratively clustered these codes into cohesive themes through iterative discussion, examining how each system supported or challenged participants’ values in prayer.

\begin{table}
  \caption{Participant Details for Design Workbook Study and Their Willingness to Use Proposed Systems}
\Description[table]{Twenty-one participants (P1–P21) with ages ranging from 20 to 39 and a mix of male and female participants. The table indicates willingness to use four AI prayer systems, with totals showing 14 for Daily Thankful, 13 for Prayer of the Past, 9 for Question for Your Prayer, and 12 for Prayer of the Other.}
  \small
  \label{tab:participants_phase2}
  \centering
  \renewcommand{\arraystretch}{1.15}
  \setlength{\tabcolsep}{3pt}
  \rowcolors{3}{gray!15}{white}
  \newcolumntype{Y}{>{\centering\arraybackslash}X}
  
  \begin{tabularx}{\columnwidth}{>{\centering\arraybackslash}p{0.09\columnwidth} >{\centering\arraybackslash}p{0.09\columnwidth} >{\centering\arraybackslash}p{0.13\columnwidth} Y Y Y Y}
    \toprule
    \rowcolor{white}
     &  &  & \multicolumn{4}{c}{\textbf{Willing to Use the System}} \\
    \cmidrule(lr){4-7}
    \rowcolor{white}
    & \textbf{Age} & \textbf{Gender} & \makecell{\textbf{Daily}\\\textbf{Thankful}} & \makecell{\textbf{Prayer of}\\\textbf{the Past}} & \makecell{\textbf{Question}\\\textbf{for Your }\\\textbf{Prayer}} & \makecell{\textbf{Prayer of}\\\textbf{the Other}} \\
    \midrule
    P1  & 28 & Male   & O & O & O & \\
    P2  & 24 & Female & O &   &   & O \\
    P3  & 25 & Male   & O &   &   & O \\
    P4  & 27 & Female & O & O &   & \\
    P5  & 22 & Female & O &   & O & O \\
    P6  & 26 & Male   & O & O & O & \\
    P7  & 25 & Male   &   & O &   & O \\
    P8  & 24 & Female & O & O &   & \\
    P9  & 24 & Male   &   & O & O & O \\
    P10 & 26 & Female &   & O &   & O \\
    P11 & 29 & Male   & O &   &   & \\
    P12 & 23 & Female &   & O & O & O \\
    P13 & 20 & Female & O & O & O & \\
    P14 & 28 & Female & O &   &   & O \\
    P15 & 30 & Female & O &   & O & \\
    P16 & 26 & Male   &   & O &   & \\
    P17 & 25 & Female & O & O & O & O \\
    P18 & 39 & Male   & O & O &   & O \\
    P19 & 34 & Male   &   &   &   & O \\
    P20 & 27 & Male   & O &   & O & \\
    P21 & 27 & Female &   & O &   & O \\
    \midrule
    \rowcolor{white}
    \textbf{Total} &  &  & \textbf{14} & \textbf{13} & \textbf{9} & \textbf{12} \\
    \bottomrule
  \end{tabularx}
\end{table}

\subsection{Design}

In this section, we describe how the four design concepts were developed. We first present the design process transparently, followed by detailed descriptions and the design rationale for each concept. Our initial design approach was to create one design for each of the four values identified in the previous phase, leading to four concepts. In doing so, we aimed to augment the practices our diary study participants engaged in to realize their values. Based on several iterations between the first and second authors, we came up with the following concepts:

\begin{itemize}

\item For \textit{self-reflection}, we designed \textit{\textbf{Daily Thankful}}, a service that curates moments of gratitude from users’ digital footprints to support daily reflection.

\item For \textit{serenity}, we designed \textit{\textbf{Prayer of the Past}}, a service that resurfaces past prayers in a seemingly random manner to evoke a sense of response.

\item For \textit{authenticity}, we designed \textit{\textbf{Question for Your Prayer}}, a conversational service that supports deeper reflection through gentle, question-based prompts without mediating the act of prayer.

\item For \textit{community}, we designed \textit{\textbf{Prayer of the Other}}, a service that presents users with others’ prayers with similar themes to broaden their experience.

\end{itemize}

However, in discussing these concepts, we realized that these systems do not exclusively support a single value. For example, while \textit{Question for Your Prayer} was designed to support authenticity, engaging in dialogue about one’s prayer with the agent can also facilitate self-reflection. For this reason, rather than presenting each system as targeting a specific value, we instead discuss how each design may relate to multiple values in the following section.

In particular, while discussing the concepts among the authors, the value of authenticity emerged as an overarching consideration across all design concepts, rather than a value that could be directly augmented or supported. This was because, as participants discussed in the previous phase, the very presence of technology could at times disrupt the prayer experience, resonating deeply with \citeauthor{smith2026spirit}'s argument that spiritual design must deeply consider the degree of technological presence \cite{smith2026spirit}. 

To address this, we approached authenticity as an ever-present value in spiritual AI systems, rather than something to design \textit{for}. We argue that authenticity is shaped by the degree of \textit{agency} that AI takes in directing and shaping the prayer experience. Here, we operationalize agency as \textit{the degree to which an entity has a say in shaping the interaction}. For instance, in some concepts, the AI agent may take a more active role in guiding users in a particular direction while praying. In such cases, we interpret the AI agent as exerting greater agency in guiding users through prayer, while users’ agency in shaping the interaction is correspondingly reduced. In contrast, some concepts may position the AI agent in a subtler role, gently “nudging” users without being overtly visible or by allowing greater interpretive space. In these cases, users retain greater ability to interpret and shape their own experience. Here, the AI agent’s agency is smaller, while the user’s agency is greater within the interaction. Connecting this notion back to the value of authenticity, we speculated that granting users bigger agency would be interpreted as a more authentic spiritual experience, as it would effectively reduce the technology’s (felt) presence. 

Our idea of agency between the user and the AI agent is informed and inspired by \citeauthor{bardzell2015user}’s work on conceptualizing the \textit{users} as ``subjectivities of information.'' In their work, \citeauthor{bardzell2015user} introduce the notion of \textit{subject} (“structures that people are thrust into”) and \textit{subjectivity} (“the felt experience and creative agency of individuals within that situation”). Through this distinction, they invite HCI and design scholars to attend to what users \textit{can do} (subjectivity) within the systems they are situated in \cite{bardzell2015user}. Building on this perspective, our notion of agency concerns the extent to which a design of an AI agent affords users the capacity to exercise their subjectivities.

With this framing, we detail each concept below with particular attention to (1) how each design was inspired by and engages with the values identified in our study, (2) its feasibility through specific AI mechanisms, and (3) the degree of agency dynamics we speculate between the AI agent and the user.

\subsubsection{Daily Thankful}

collects users' digital footprints—instant messages, social media posts, emails, and other activities—within the boundaries that users set. At day's end, the system compiles an inventory of noteworthy events for which users can express gratitude, ranging from heartwarming messages from friends and family to the beautiful sunset they captured. This compilation serves as a resource for users to draw on to engage in prayerful acts of thanksgiving.

The concept was inspired by users actively recording their daily lives to find moments to pray for. By providing them with specific moments and instances to reflect on, we aimed to enhance the value of self-reflection. The concept aimed to provoke a more exhaustive reflective experience, especially highlighting how they kept track of what to be thankful for that day.

Here, the AI agent’s role in the system is to identify meaningful moments from users’ daily lives and generate curated prompts for gratitude. To do so, the system draws on a range of digital footprints, including textual and visual data. Specifically, the agent employs computer vision techniques (e.g., object and scene detection) to interpret images, and natural language processing (NLP) techniques, such as large language models (LLMs), to analyze textual data. Through multimodal aggregation, the agent integrates heterogeneous data sources, synthesizing and curating them into a coherent, presentable form.

Regarding authenticity and the relationship between users and AI agencies, \textit{Daily Thankful}’s AI system takes a more active role by collecting users’ personal data and curating prayer guidelines, while users are placed in a more receptive position, receiving these suggestions. Nevertheless, they retain a degree of agency by choosing which of the suggested topics to focus on during prayer, thereby integrating the system’s guidance in ways that feel personally meaningful.

\subsubsection{Prayer of the Past}

is an agent designed for users who maintain daily prayer journals. At seemingly random intervals, the system surfaces a past journal entry. While this resurfacing appears random, the AI agent accounts for the user’s recent state (e.g., current concerns) to select entries that may be meaningful or supportive. When writing their prayer journal for the day, users can reflect on and provide feedback regarding the resurfaced entry. This feedback loop enables the system to better understand which types of journal selections users find meaningful, thereby refining future selections.

The concept was inspired by (1) how users delve into past prayer journals to reflect on their personal and spiritual growth, (2) how they actively interpret random daily encounters as responses to their prayers, and (3) the slow design artifact \textit{Olly} by \citeauthor{odom2019investigating}, which prompts reflection by randomly surfacing past data with minimal user control \cite{odom2019investigating}. \textit{Prayer of the Past} primarily aims to support the values of self-reflection and serenity. By allowing users to revisit their past prayers and compare them with their present selves, the system encourages a more engaged, reflective experience. Additionally, by introducing an element of randomness in the printing of past journals (in both content and timing), the system creates opportunities for interpretation, such as “Why was this particular journal presented to me at this moment?”, closely mirroring how users found serenity by making meaning from serendipitous occasions interpreted as divine.

Here, the AI agent’s role is to retrieve and present past prayer journal entries that resonate with the user’s current state. To do so, the system draws on users’ historical journal data, along with signals of their recent concerns (e.g., recent entries or inputs). Specifically, the agent employs NLP techniques such as LLMs to encode and analyze both past and recent journal texts (e.g., topic modeling, sentiment analysis, or embedding-based similarity). These representations are then used in a retrieval system (e.g., semantic search) to identify entries that are contextually relevant rather than randomly selected.

Regarding the value of authenticity, we considered this concept to embody minimal AI presence. Interaction occurs mainly through traditional means, such as keyboard typing and physical printing, with the AI never presenting itself as an agent that directs how or why users should reflect on a particular prayer. Instead, it simply resurfaces past entries without guidance or modification, granting users interpretive agency over the printed journals. At the same time, the fact that users do not know how or why these journals were printed, and have no direct control over the serendipitous moment of encounter, can also be seen as diminishing their sense of agency.

\subsubsection{Question for Your Prayer}

is a conversational AI system designed for users seeking to deepen their reflective prayer experience. When a user feels their prayer is superficial or lacks earnestness, they can enter the content of their prayer into the system. The AI then engages the user with a series of probing questions to help them delve deeper into the essence of their prayer. Sample inquiries might include: ``What prompted those feelings?'', ``How could you improve the situation?’', or ``This particular Bible verse discusses scenarios requiring increased effort. How do you perceive its relevance to your circumstances?''

This concept was inspired by how users actively sought to deepen their prayer while retaining their own agency, rather than being told how to pray, similar to how P7 from the diary study appreciated a Bible application that posed reflective questions, rather than ChatGPT writing his prayers for him. We also considered how such questions could prompt users to confront the emotional difficulty of disclosing their wrongdoing or doubts about their beliefs, an aspect many participants regarded as essential to reflective prayer. In doing so, \textit{Question for Your Prayer} aims to support a more honest and sincere prayer experience, closely aligning with the value of self-reflection by fostering and encouraging deeper introspection.

Here, the AI agent’s role is to facilitate deeper reflection by generating contextually relevant, probing questions based on the user’s prayer content. To do so, the system employs NLP techniques such as LLMs to parse and interpret the input prayer, identifying key themes, emotions, and underlying concerns. The LLM then generates tailored, open-ended prompts that encourage introspection, ensuring relevance and depth based on the user's specific context.

Regarding the value of authenticity, the AI retains a limited degree of agency by guiding, provoking, and encouraging deeper reflection. Yet it does not reflect on behalf of users or steer them toward a predetermined direction, allowing users to retain agency in shaping their own prayer experiences.

\subsubsection{Prayer of the Other}

displays the prayer journals of fellow individuals whose entries relate to the user’s own. When a user shares their prayer journal, the AI identifies related prayers—those similar in topic, that expand on what the user wrote, or that offer responses to what the user prayed for—and presents them to the user.

This concept was inspired by moments when participants found value in praying with others, such as discovering people with similar concerns, gaining insight into social issues they had been unaware of, or learning new prayer practices. \textit{Prayer of the Other} is connected to the value of community, as it allows users to enrich their personal prayer experiences through others’ journals and find emotional resonance with people who share similar beliefs.

Here, the AI agent’s role is to identify and present prayer entries from others that meaningfully relate to the user’s own prayer. To do so, the system employs NLP techniques such as LLMs to encode users' prayer texts into semantic representations. These representations are then used in a similarity-based retrieval system (e.g., semantic search) to match and surface relevant entries from a shared corpus.

Regarding the value of authenticity, similar to \textit{Prayer of the Past}, the AI does not interpret or guide users in any particular direction. Instead, it simply presents others’ prayer journals, leaving users free to make sense of the results on their own terms and thereby affording a higher level of agency. At the same time, the system can also be understood as reducing users’ agency, as they have no control over which prayers are presented to them or why—an intentional design choice meant to create a sense of serendipitous encounter.

\subsection{Findings: Co-Exploring the Implications of Spiritual AI}
Based on our analysis, we present our findings, focusing on how each value was expected to be affected by the introduction of AI systems.

\subsubsection{Balancing \textit{Self-reflection} and \textit{Authenticity}}
Our findings reveal significant concerns about the AI system reflecting ``on behalf of'' users, in other words, taking the agency away from the users, which many felt undermined the true meaning of self-reflection. When discussing \textit{Daily Thankful}, participants emphasized that genuine gratitude involves appreciating not only easily recognizable blessings but also finding reasons to be thankful in challenging situations. In discussing this design, P10 shared that: \textit{``The purpose of a gratitude prayer is to quietly reflect on one’s day alone, revisiting the things one is thankful for and bringing the day to a close. In that sense, relying on AI to filter and surface all moments of gratitude from one’s digital life, rather than doing it on one’s own, feels somewhat uncomfortable’’}. In other words, the essence of a good prayer lies in the conscious effort to personally recognize and reflect on blessings, and automating this process was seen as diminishing its purpose.

This concern was particularly evident in participants’ remarks about preparing data for AI input. Excluding \textit{Daily Thankful}, three of our concepts required users to manually enter their prayer journals. While some participants found this task tedious and expressed a preference for systems that could automate or simplify the process (For instance, in discussing \textit{Prayer of the Past}, P1 shared: \textit{``It would be helpful if there were a feature that allows users to input spoken prayer directly, such as support for voice-based prayer’’}), while some viewed the preparation, curation, and creation of data itself as a meaningful reflective practice (In discussing the same design, P5 shared: \textit{``I think I would feel a sense of efficacy in having a more structured way of engaging in prayer’’}). In the latter case, the value of our proposed design was perceived as motivating users to reflect, write, and document their journals more diligently, rather than in the capabilities of the AI agent itself. 

\subsubsection{How Prayer AI Can Diminish the Value of \textit{Serenity}}
Participants identified several ways in which AI systems could harm the value of serenity. They expressed concern that the AI's unpredictability and vast data-processing capabilities could provoke unexpected negativity, ultimately disrupting their calm. 

Regarding AI’s unpredictability, participants again expressed concern about moments in which they would have little agency or control over what was presented to them. In discussing \textit{Prayer of the Other}, P14 voiced the necessity of having some level of moderation to monitor \textit{``content that mentions or criticizes specific individuals, includes profanity or explicit material, or appears to be spam.''}. For \textit{Prayer of the Past}, P13 shared \textit{``If a past prayer is tied to a negative experience, resurfacing it could bring back traumatic or unpleasant memories for the person praying.}, illustrating how having low agency and control over what they are presented with led to diminished serenity.

Additionally, participants voiced concerns about being overwhelmed by the sheer volume of data provided by AI. In the case of \textit{Daily Thankful}, participants were worried that the moments identified by the system might be too numerous to address individually. P12 shared: \textit{``It feels like the AI is picking out even the smallest things to be grateful for, and if I had to read all of that every day, I think it might get overwhelming and give me a headache,’’} while other reflecting that this overwhelming amount of data could also lead to a sense of guilt if they were unable to give thanks for all the prompts provided.

Moreover, participants reported feeling intimidated by the system, particularly when the AI assumed a dominant role in the interaction. For instance, in the \textit{Question for Your Prayer} concept, P2 felt the system was overly interrogative: \textit{``It feels a bit scary… like being in an interview where a very analytical person keeps probing with questions. Even if it responds in a more empathetic tone, it would still feel rigid, as if the AI is just producing pre-set outputs.''} 

Overall, participants felt that the value of serenity would be significantly compromised when AI enters their prayer experience, particularly due to the overwhelming presence of unpredictable, unwanted, and difficult-to-process outputs. These responses were also understood as moments where users have little control over what they encounter while interacting with AI agents. Such a loss of agency in managing a private and serene practice like prayer was expected to cause considerable distress.

\subsubsection{Concerns Regarding \textit{Authenticity}}
Regarding the value of authenticity, participants exhibited conflicting reactions to using technology for religious purposes, echoing findings from our diary study. While some appreciated the perceived ``impartiality'' of AI compared to guides and supports from humans (often considered ``partial''), many expressed discomfort with the idea of technology intervening in such a sacred realm.

On a positive note, the non-human nature of AI contributed to its image as an impartial, effective, and private prayer assistant. For instance, when discussing \textit{Question for Your Prayer}, P5 stated that she would \textit{``share things that are difficult to open up about to other people,''} anticipating greater honesty in their prayers, leading to a more authentic experience. This was because AI, being detached from their social circle and human values, could provide guidance without threatening their social image. Moreover, with the presumed impartiality of \textit{Question for Your Prayer}, participants appreciated the AI's ability to offer new perspectives and suggest Bible verses relevant to their situations, helping align their prayers more closely with divine will.

In some cases, the image of AI as impartial was reinforced. For example, while discussing \textit{Daily Thankful}, P6 shared that \textit{``adding a religious meaning made the AI’s observation of their personal life feel less intrusive''} and did not feel like a privacy violation. In these moments, the system was perceived as being used for ``goodness,'' which led to a sense of comfort with AI systems observing their daily lives. 

However, many participants questioned this very feature, expressing doubts about whether AI can truly remain impartial. P4 worried about the AI’s ability to provide appropriate and ethical responses to controversial issues that even humans continue to grapple with, stating: \textit{``If the interaction turns into a form of debate with AI, it may position the system between the user and God.''} P15 also said \textit{``AI-generated content could be quite sensitive, as it has the potential to influence the direction of one’s prayer. Even small shifts could open up concerns around misleading or cult-like influences,''} raising concerns that AI could generate misguided or extremist interpretations of religious texts, potentially leading to distorted or harmful spiritual experiences. 

Even if AI \textit{could} somehow achieve a state of providing neutral guidance, participants still felt that the presence of technology in their intimate relationship with the divine was disturbing, echoing our previous findings. P21 shared that, \textit{``prayer is something done alone with God. If AI is used, it may shift from prayer to seeking advice or a conversation partner that is not God,''} sharing her discomfort with entrusting spiritual experiences in AI, feeling that prayer is a personal dialogue with the divine that should not be mediated by another entity. This discomfort was further compounded by the notion of communicating with an intangible entity, as seen in the conversational form of \textit{Question for Your Prayer}; P7 reflected: \textit{``I don’t really find meaning in talking to AI, so I don’t think I would want to engage with it seriously. It feels like talking to something intangible and without real presence. In that case, I would rather just talk to God and reflect on my own.''} These remarks may showcase the inherent limits that these agents could, since what matters, in essence, is \textit{``whether there is a human or a machine on the other side of the screen.'' }(P9)

\section{Discussions}
In our study, we identified four key values related to prayer: self-reflection, serenity, community, and authenticity. Drawing on these values, we designed four conceptual spiritual AI systems, which participants were invited to discuss and examine to identify diverse moments where these values were both enhanced and undermined. In the following sections, we delve into the implications of our findings for the design of spiritual AI systems, alongside broader considerations in human-AI interaction, particularly concerning AI non-use, user agency, and the potential benefits of the inexplicability of AI systems.

\subsection{Balancing AI and User Agency for Authentic Experience}

In our study, one of the most significant values shaped by the introduction of AI in spirituality was authenticity. During the diary study, participants emphasized a “sense of connection” with the divine as central to their spiritual experience and questioned whether technological intervention might undermine this authenticity. Moments where authenticity was diminished were discussed more thoroughly in the design workbook study, particularly during conversations about \textit{Daily Thankful}. The reflective journey that users undertook in preparing for gratitude prayer, such as actively identifying, acknowledging, and recording moments of gratitude, was itself a meaning-making process, as revealed in the diary study. However, in the case of \textit{Daily Thankful}, when the system identified moments of gratitude on behalf of users, participants felt that the essence of prayer was diminished. 

Existing research has examined how technological automation can undermine the meaning-making process of spiritual practices~\cite{macwilliams2013virtual, helland2007diaspora}. These studies highlight how virtual pilgrimages diminish the significance of the experience, as its core meaning often lies in the arduous physical journey and the practitioner's active participation. Recent works in spiritual technologies have also highlighted how the framing for such technologies should move beyond the `outcome' of spiritual practices to recognize and foreground the `process' of them and how they create meaningful experiences \cite{mim2026making,zhu2026toward}. In other words, users do not expect technology to ``live on their behalf'' for convenience's sake. This reflection underscores the importance of carefully balancing agency and efficiency in AI usage: preserving users' agency in the meaning-making process to ensure that they retain an authentic experience, even while benefiting from technological assistance.

In this context, we expand our discussion along two threads. First, we examine instances where the presence of technology diminished spiritual experiences and consider the implications of spiritual technologies in relation to non-use. Second, we reflect on moments in our study where users’ agency was preserved, enabling them to feel authentic in their prayer practices while still benefiting from AI’s capabilities, providing further implications for designing AI systems in nuanced and sensitive human experiences.

\subsubsection{Recognizing the Value of Inconvenience}
As discussed, participants frequently expressed discomfort with the presence of technology in their spiritual activities. For example, they felt the value of self-reflection was diminished when the AI in \textit{Daily Thankful} assumed agency over their reflections on daily life. Likewise, the value of serenity was perceived as threatened in \textit{Question for Your Prayer}, where participants felt the AI imposed moral judgments through its prompts, provoking significant unease. These accounts resonate with findings from \citeauthor{xygkou2023conversation}, who show that many individuals regard chatbot-like agents in intimate contexts as inferior to human connections, preferring meaningful interactions with friends and family in line with the value of community~\cite{xygkou2023conversation}. Above all, participants voiced concern over introducing AI, a technology whose biases, capabilities, and risks remain opaque even to its own developers, into sacred and intimate relationships with the divine, thereby undermining the authenticity of their spiritual experiences.

These concerns raise a fundamental question: Should AI be introduced into the realm of spirituality at all? This question underscores the importance of discussing the concept of non-use in both AI and techno-spirituality. Studies on non-use in design and HCI explore why and how individuals deliberately choose not to engage with specific technologies~\cite{baumer2013limiting, baumer2014refusing}. By paying special attention to such moments, HCI and design scholars avoid their research becoming overly techno-solutionist and critically assess the socio-cultural context surrounding the notions of the said technology to envision, design, and implement designs that better cater to people's realities \cite{baumer2011implication}. This conversation is particularly critical in an era where technology, including AI, increasingly permeates everyday life. If we consider cases of individuals resisting the incorporation of AI into religious or spiritual practices, the intentional decision to preserve certain aspects of life free from AI becomes an invaluable area of exploration in human-AI interaction.

The discussion of non-use has been examined across various contexts relevant to our study. Researchers in human-AI interaction have explored instances where individuals actively resist AI systems in workplaces~\cite{boucher2024resistance,cha2025understanding}. Similarly, studies on spiritual technologies have examined cases where people opt out of using religious technologies, such as the Amish rejection of certain technologies~\cite{ems2014ict, woodruff2007sabbath, macwilliams2013virtual, helland2007diaspora}, and the implications of such decisions for preserving value-laden, ambiguous transcendent experiences such as meditation~\cite{markum2020digital}. Given these perspectives, we caution against approaches that aim to use AI in spiritual practices to "solve" non-existent problems or propose improvements disconnected from the values and situated realities of people~\cite{ferdous2017celebratory, smith2024un, baumer2011implication}. 

This, however, does not suggest refraining entirely from developing spiritual AI technologies. Rather, the conversation on non-use emphasizes its layered nature: a constant (re)negotiation of disengagement with technology, where the degree of disengagement varies across contexts~\cite {baumer2015importance}. For instance, participants in our diary study reflected on how Christian believers’ perspectives on online worship evolved post-pandemic. Rather, we might consider this provocation to be the understanding that preserving a certain level of inconvenience should be recognized, understood, and considered in creating AI systems, recognizing that ``making things easy'' may not always be the goal of technological aids~\citep{nilsson2020visions,cox2016design}. As our relationship with AI and its role in spiritual contexts continues to evolve, we emphasize the importance of resisting the assumption that the ``use" of technology is the default. Instead, we advocate actively considering and adopting non-use as an essential design space for AI technologies in human contexts deeply embedded with nuanced, ambiguous values, including spirituality.

\subsubsection{AI as a Catalyst rather than a Problem-Solver}
Expanding on the previous discussions, we also consider how spiritual AI \textit{may} respect the value of authenticity while still providing technical efficacy. One possible pathway is to position AI as a catalyst or provocateur rather than a solution provider. In the case of \textit{Question for Your Prayer}, users appreciated interacting with the system because it encouraged deeper reflection through questions rather than providing direct answers. In this way, users retained control over shaping the meaning and experience of their prayer, with the AI serving only to suggest new topics or directions for exploration. Here, allowing users the freedom to accept or dismiss the AI’s suggestions, positioning the AI as a catalyst rather than a directive guide, was expected to enrich the reflection process without displacing the spiritual experience. 

Understanding how AI can preserve and respect users' agency is crucial not only for designing spiritual AI systems but also for comprehending the broader human-AI relationship. As AI increasingly permeates our daily lives across diverse realms, a key question arises: How will design and HCI researchers define, understand, and study human experiences facilitated or enabled by AI? In our study, we highlight how users value living an authentic spiritual life, maintaining their agency in shaping their realities. 

This desire for authenticity likely extends beyond spirituality. For example, a painter paints not just to create a finished piece, but to find meaning in the process itself. The frustration when things don’t go as planned, the joy when they do, and the excitement and nervousness of presenting their work to others are integral parts of imbuing the experience and the outcome with meaning. If a generative AI instantly produces the desired painting, the user's agency in the creative process is diminished, leading to a less authentic experience. It is how we experience with the agency, without having someone or something live on our behalf or telling us how to live. 

These reflections deeply resonate with the concept of bidirectional AI alignment suggested by \citeauthor{shen2024towards}, which focuses not only on how AI aligns with human values but also how human behavior is influenced by AI~\cite{shen2024towards}. Thus, we propose that design and HCI researchers actively recognize and discuss the meanings of human autonomy and agency in the age of AI, and consider how these values and discussions may be integrated into our vision for the future of human-AI relationships \cite{kwon2026ai}.

\subsection{Inexplicability of AI as a Room for Interpretation, Boosting Serenity}

Due to the complexity of AI algorithms and the vast amount of data they process, it is often difficult for humans, even for AI developers, to fully understand how these systems operate~\cite{rudin2019stop}. This opaque nature of AI, often described as a ``black box'' or lack of transparency~\cite{rudin2019stop}, has long been identified as a limitation, prompting many researchers to propose solutions under the growing agenda of ``explainable AI''~\cite{dwivedi2023explainable}.

Interestingly, however, in a spiritual context, we found that this very obscurity can provide users with a form of efficacy. The inability to fully explain how AI works leaves room for interpretation, allowing users to engage with its outputs in ways that enhance their sense of serenity. In our design workbook study, concepts such as \textit{Prayer of the Past} and \textit{Prayer of the Other} involved the AI presenting users with past or other people’s prayer journals. In discussing what they liked about each system, participants noted that \textit{Prayer of the Past} might give them the chance to revisit moments of gratitude or to recognize their growth and resilience by reflecting on past journals that describe hardships. With \textit{Prayer of the Other}, they found comfort in reading how others prayed, drawing strength from shared struggles, gaining fresh perspectives, and even identifying new social issues they could include in their prayers.

We interpret these moments as demonstrations of users’ capacity to shape their own reality with the information provided, which deeply resonates with the notion of designs that allow users' agency and subjectivities \cite{bardzell2015user}. In other words, participants anticipated that whatever content the system offered, they could reinterpret it to fit their present circumstances. This reflects how users reified serenity in our diary study, where peace was found by treating random events as divine responses. In essence, participants discovered serendipity in the AI’s unexpected outputs, interpreting them in ways that brought joy, peace, and a sense of efficacy.

Such recognition of unexpectedness and ambiguity as serendipitous experience appears across HCI \cite{odom2019investigating, jang2025journey, kim2022slide2remember,jiang2021supporting} and techno-spirituality research \cite{wolf2023designing, woodruff2007sabbath}. For example, \citeauthor{woodruff2007sabbath} examined the Jewish practice of Sabbath, during which technology use is prohibited every Friday. They describe how some employ home automation to prepare for the Sabbath, such as programming lights to turn off at specific times. When these systems fail or behave in unintended ways, many interpret the disruptions as divine will and incorporate them into their spiritual practice. \citeauthor{woodruff2007sabbath} conceptualize this as a ``surrender of control,’' where relinquishing mastery over technology fosters peace through acceptance and reflection. We extend this notion by emphasizing that, in these moments of surrendering control, users are not simply passive recipients of technological breakdowns; rather, they actively enact their subjectivities, reinterpreting technical infrastructures in ways that affirm their agency and transform disruption into meaningful and comforting experiences \cite{bardzell2015user}.

We also observed such ``surrender of control'’ in our study. In discussing \textit{Prayer of the Past} and \textit{Prayer of the Other}, participants acknowledged their inability to determine what the AI would provide, yet they reflected on what was given and found their own version of peace. This insight suggests new directions for future AI design. While much current discourse emphasizes explainability and the reduction of so-called ``hallucinations,'' in contexts where user interpretation is central (such as spirituality, fortunetelling \cite{cho2025shamain,kwon2026ai}, or creative practices), preserving ambiguity and serendipity may foster novel forms of user satisfaction. More broadly, embracing the incomprehensible aspects of AI as spaces for diverse interpretation can help cultivate and sustain users’ agency and authenticity in human–AI interactions, as discussed in the previous section.

This implication, once again, extends beyond spirituality into other human-AI research agendas. In brainstorming, for instance, AI's role could be seen as formative rather than summative, to provoke new directions of thought rather than provide complete ideas. By sparking additional thinking in various ways, regardless of AI's intent, it may enhance creativity by leading users into new directions. In short, embracing the inexplicable and unexpected nature of AI as a meaningful design space may open novel and exciting avenues for various fields.

\subsection{Critical Reflections and Future Works}

To conclude our study, we offer critical reflections on our work and discuss how it may prompt future research. First, we reflect on the process of designing the four AI prayer support systems. While we provided a value-driven rationale for each concept, we acknowledge that our ideation and design process was not exhaustive. Our approach began with mapping each of the four identified values to a corresponding design concept, rather than systematically exploring a broader space of alternative designs and iteratively refining or selecting among them. Relatedly, although we introduce agency as a conceptual lens to reason about authenticity across designs, we did not employ a formalized design framework or systematically vary design parameters to explore the full range of possible configurations of agency.

Further, the discussion points we raise on how we might leverage the non-use, opacity, serendipity, or ambiguity of AI agents, as well as the notion of agency as an opportunity for AI technologies, remain at the level of broad ideas rather than concrete design guidelines. That said, we look forward to future work that explores these concepts more concretely through rigorous design studies.

We also reflect on the scope of our study. Our work is deeply situated in the experiences of younger South Korean Christians. This means that our participants represent a very particular context: they are familiar with and adept at using various technologies in their daily lives while also navigating a secular society alongside those with no or different beliefs, with varying degrees of religious importance in their lives. That said, we clearly set these boundaries and reiterate that the goal of this study lies less in achieving generalizability, and more in provoking discussion around what ‘could be.’ We look forward to future work that similarly examines local uses and expectations of technologies in spiritual and religious contexts, as such experiences and expectations vary significantly across cultures.

\section{Conclusion}
In our study, we examined how envisioning spiritual AI systems that aid prayer experience could spark meaningful conversations in human-AI interaction, especially in the (non-) design of AI agents that respect human agency. To conclude, we invite design and HCI researchers to envision AI systems that not only leverage computational capabilities but also empower users to exercise agency and cultivate authentic human experiences.

\bibliographystyle{ACM-Reference-Format}
\bibliography{01reference}

\end{document}